# *Potential Biased Outcomes on Child Welfare and Racial Minorities in New Zealand using Predictive Models: An Initial Review on Mitigation Approaches.*

Sahar Barmomanesh[1a], Victor Miranda-Soberanis[a]

[a] Department of Mathematical Sciences, Auckland University of Technology, 6 St Paul Street, Auckland City, Auckland, 1010, New Zealand

*Abstract*— Increasingly, the combination of clinical judgment and predictive risk modelling have been assisting social workers to segregate children at risk of maltreatment and recommend potential interventions of authorities. A critical concern among governments and research communities worldwide is that misinterpretations due to poor modelling techniques will often result in biased outcomes for people with certain characteristics (e.g., race, socioeconomic status). In the New Zealand care and protection system, the over-representation of Māori[2] might be incidentally intensified by predictive risk models leading to possible cycles of bias towards Māori, ending disadvantaged or discriminated against, in decision-making policies. Ensuring these models can identify the risk as accurately as possible and do not unintentionally add to an over-representation of Māori becomes a crucial matter. In this article we address this concern with the application of predictive risk modelling in the New Zealand care and protection system. We study potential factors that might impact the accuracy and fairness of such statistical models along with possible approaches for improvement.

## INTRODUCTION

Decision making tasks are regularly supported by predictive models developed by machine learning algorithms on historical data. Some examples are credit scoring, college admissions, and hospital admissions [1-3]. Although predictive models have demonstrated efficiency in a variety of settings (e.g., insurance, Finance, Health), they have only recently been applied to the classification of risk in Child Protective Services [71-73]. The number of child welfare agencies developing and deploying predictive tools that use government data and machine learning algorithms to predict child maltreatment is increasing. For instance, several government agencies in the US are currently testing and incorporating predictive risk models into their child welfare systems [4-6]. Three of the most well-recognized predictive tools currently in use by child welfare agencies in the US are the Eckerd Rapid Safety Feedback tool (ERSF), developed in Florida [4], The Allegheny Family Screening Tool (AFST), developed by and used in Allegheny County, Pennsylvania [5], and the Douglas County Decision Aid model (DCDA) stablished specifically by the Douglas County in Colorado [6]. As part of its early intervention and prevention efforts, New Zealand also has considered the use of predictive risk models to assist professionals in identifying children at risk of abuse or neglect [7]. A number of predictive risk models were developed to assess the technical feasibility and predictive validity of the New Zealand government proposal [8-10]. Based on the findings of these studies, technical feasibility studies [11], and ethical reviews [12, 13], predictive risk models based on administrative data could potentially be used to identify high-risk New Zealand children who need preventive services. However, predictive tools should be considered as complementary to professional judgment rather than as a substitute and should not be considered the sole method for identifying high risk children. While the potential value of predictive tools in decision-making is acknowledged, research also warns against ethical risks as well as risk associated with compounding surveillance bias [12, 14-16]. Specifically, in New Zealand, the risks include the potential discriminatory effects on Māori. The *feasibility study* conducted by the Ministry of Social Development [11] with focus on predictive risk models to identify new-born children in need of prevention services, reported that models developed for New Zealand population would refer children in numbers that do not always reflect their share of known maltreatment cases. In this line, Wilson, Tumen [9] found that the model developed identifies far too many high-risk children who are known to be Māori, compared to their true incidence rate of maltreatment. Among children with the highest risk scores (Top 5%), 69% were Māori, while only 61% of those known to have been maltreated by the age of Two were Māori. Furthermore, Rea and Erasmus [10] investigated the disparity between current referral rates and rates produced by the model amongst different ethnic groups. Results showed that the model referred a greater proportion of Māori children to the site than current practice, and this unusual referral rate remains unknown, so further investigation is required to determine the model feasibility in the future and so be able to identify risk as accurately as possible and to avoid unintentional bias or unfairness toward racial groups such as Māori. This step would provide a clearer platform towards an implementation phase in New Zealand. In line with this, this article aims to provide a theoretical baseline for the research on fairness-aware machine learning in child welfare settings with a focus on racial minorities in New Zealand such as Māori.

The outline of this article is as follows: Section 1 talks about unfairness and explains how bias would creep into the process of predictive modelling and consequently lead to making biased or unfair decisions. Section 2 outlines the factors that might influence the accuracy and fairness of predictive risk models developed for use by child protection services, Section 3 and Section 4 provide a review of fairness-aware machine earning and some relevant notions of fairness. Finally, Section 5 discusses an initial potential approach to bias mitigation during the process of machine learning.

### 1. BIAS OR UNFAIRNESS IN THE PREDICTIVE MODELLING PROCESS

The use of predictive models developed through machine learning algorithms is becoming increasingly prevalent to assist, or sometimes even replace human decision-makers. Evidence, however, suggest that decisions made by algorithms that are incorrectly trained can result is disproportionally unfavourable outcomes for certain groups of people in ways that resemble discrimination [17]. A study by Larson, Surya [18] argued that the Correctional Offender Management Profiling for Alternative Sanctions" tool (COMPAS), which is used in several courts across the United States to predict recidivism, was unfair to African American defendants. It transpires that the model falsely labels African American defendants as high risk at a greater rate compared to White defendants. There have also been several cases in which predictive models have inadvertently discriminated against certain social groups [19-21]. Discrimination refers to the unfair treatment of an individual based on their membership in a certain group rather than individual merits [22]. This term is ethically, legally, and socially improper and might lead to conflicts among groups. Laws at both the national and international levels prohibit discrimination on a wide range of grounds and in a wide range of settings. Accordingly, the New Zealand human rights act 1993 aims to ensure that people in New Zealand are treated fairly and, in accordance with United Nation agreements. It applies to nearly all government and private organizations. Decisions made by predictive models are also subject to New Zealand's anti-discrimination [23].

---

[1] Corresponding author (e-mail: sahar.barmomanesh@ aut.ac.nz).

[2] Māori are the indigenous Polynesian people of mainland New Zealand. Māori originated with settlers from East Polynesia, who arrived in New Zealand between roughly 1320 and 1350.

As predictive models are developed using historical data, the characteristics of the data used in the training process can significantly impact the lessons learned by the algorithm [17]. If an algorithm is trained using biased historical data, this bias can propagate through the modelling process and consequently result in biased decisions.

As child protection authorities continue to implement predictive analytic tools, questions about their reliability and validity arise. In particular, there have been concerns about the accuracy of the predictive models owing to the potential presence of errors in the data as well as prediction errors [15]. There has also been ethical concern about whether predictive analytics methods will worsen existing racial disparities in child protective services. Specifically, past studies suggest that persistent racial bias reflected in administrative data may increase error rates as well as cause discrimination or unfairness against certain groups [24]. Despite this possibility inherent in predictive risk models, supporters have emphasized that these models offer a means of tracking disparities and correcting them, which can be difficult to achieve with clinical approaches. Identifying the factors that influence the accuracy and fairness of predictive models developed for use by child welfare agencies is, however, key to improving their accuracy and correcting disparities. Therefore, some of these factors will be discussed in the next section.

## 2. Factors affecting the accuracy and fairness

During the predictive modelling process, some factors may lead to disproportionately adverse outcomes for certain groups. Several factors or mechanisms have been identified as contributing to a reduction in the accuracy or bias of child welfare predictive risk models, including characteristics of the training data, the definition of the outcome variable, and the selection of features. These mechanisms will be reviewed in turn in the following sub-sections.

### 2.1. Training data

Any predictive model is developed based on the data upon which it is trained; therefore, if the data contains errors, these errors will be reflected in the model's output [17]. Specifically, if an error changes an important characteristic of a child, the accuracy of the model for that child will be reduced [15]. In the case of child welfare predictive tools, there are reasons to be concerned about the level of error present in the data being fed into the algorithm [10, 15]. As is well known, data in government administrative systems is originally entered by humans, which makes them susceptible to human error. In some cases, names, addresses, or other vital information may be inaccurate, or it may be mistakenly linked to information from another individual.

A good example of this problem is Illinois' Department of Children and Family Services' experiment with Eckerd Rapid Safety Feedback. Eckerd reported that following the implementation of the tool for three years, repeat abuse of children placed under child welfare supervision decreased from 7.9% to 5.5%. Additionally, Eckerd has confirmed that no child abuse-related fatalities have occurred since then in Eckerd's service area [25]. Despite this, Illinois' Department of Children and Family Services terminated the trial in December 2017 due to the tool's inaccuracy following the deaths of two young children within a month of each other. As later revealed, neither of those children was classified as high risk by the model and based on a review by Illinois officials, the inaccurate predictions were caused by an error in the data fed into the model [14, 15]. As a result, it is imperative to carefully link information about individuals across a variety of databases. The feasibility studies and ethical reviews on the use of these models have concluded that, although the linkage of administrative data is feasible, it is susceptible to error [11, 12]. Therefore, a system of evaluation would be required.

Furthermore, if machine learning algorithms learn from the data in which prejudice has played some role in the past, that rule might replicate prejudice involved in those earlier cases [17, 26, 27]. As an example, if learning algorithms are trained based on social workers' past decisions, which showed racial discrimination towards a particular race, the new model will also exhibit discriminatory behaviour toward that group [28]. There is therefore a risk that predictive modelling approaches can magnify bias in data and in risk assessments made by child welfare professionals.

Biased outcomes may also occur in cases where the data is unbiased but is not well-sampled [29]. For instance, overrepresentation of a certain group in the data set can lead to disproportionate adverse effects on those groups [17]. Examples are given in [14, 15, 28]. Particularly, in New Zealand, the over-representation of Māori in the care and protection system has been known for some time and that can be a reason for potential discriminatory effects on Māori. Official statistics and recent quantitative exploration of disparities for Māori children in the care and protection system acknowledge the existence of disparities between the involvement of Māori children and children of New Zealand European and Other ethnicities with the care and protection system [30]. As a result, predictive risk models are at risk of intensifying the overrepresentation of Māori, and therefore their use in decision-making could encourage a cycle of bias that, in turn, could lead to the disadvantage or discrimination of Māori [10]. There is evidence to suggest that these sources of bias can be controlled more effectively by using a model that is more accurate and carefully developed [11, 12].

## 2.2. Definition of the Outcome Variable

In predictive modelling, the outcome variable is used to define the outcome that the developer is seeking. Models cannot predict an outcome successfully unless the outcome variable is well defined and correlates well with the ground truth. A different choice for the outcome variable may have a greater or lesser adverse impact on some groups than others. As a result, there are concerns about entering bias at this stage in the modelling process [15, 17]. Take as an example, in version one of the Allegheny Family Screening Tool [31], the models were developed to predict two outcomes: re-referral and placement. However, it was revealed later that the re-referral model gives a high score to children involved in custody disputes or other circumstances where there are regular calls about the same issue. Consequently, it did not appear to be strongly correlated with outcome measures such as concern, serious abuse, and neglect. As well as this, racial bias or prejudice was embedded in the initial referrals. Therefore, a model predicting future referrals tended to overrepresent black children compared to white children. Accordingly, the second version of the Allegheny Family Screening Tool was limited to predicting significant safety issues that usually result in a court order for placement. Considering the New Zealand population, findings from the study by [10] showed that the model would refer more Māori children and young people to the site than under current practice. Therefore, future work should consider the over-representation of these groups in the care and protection system and ensure the outcome variable is defined in such a way that does not reflect an over-representation that is inconsistent with their share of actual risk.

## 2.3. Variable selection

Variable selection is the process by which developers decide which variables to use in their analysis. These decisions can have significant consequences for the treatment of certain groups if the factors which are a better representation of these groups are not well represented in the set of selected variables. This may lead to systematically less accurate classifications or predictions about those members. A possible explanation is that the absence of that variable, regardless of its significance for the outcome, will cause other correlated variables to take on a weight that does not explain the influence of the missing variable while obscures its significance [15]. Additionally, the information necessary to achieve accurate outcomes might exist at a level of detail that the selected features fail to achieve or uncover critical points of disparity [17]. The use of ethnicity as a predictor variable has been controversial due to concerns regarding racial stereotypes and the allocation of interventions according to race. Considering this, New Zealand studies have not included race as a predictor variable but tested the final model against different ethnic groups using it [8-10]. However, this idea of 'fairness through unawareness' is ineffective due to the existence of correlated variables with the sensitive variable [32]. That means even if a developer chooses not to use this variable, it may be reflected in other variables. So, the developer is indirectly using this feature encoded in other attributes. This is also known as the 'redlining effect' in the literature and often leads to indirect discrimination [33]. The use of zip codes as a predictor is one example. It is known that some suburbs may have a higher proportion of specific racial groups than others, thus providing a strong correlation between race and zip code. In addition, due to institutionalized racial bias (e.g., criminal justice history), there may exist other highly correlated predictors with race that suggest race is still a significant factor [31]. Zliobaite [34] advises that the inclusion of protected characteristics such as *race* are needed in the model development process to actively ensure the resulting model is fair. Section 3 provides a review of the approaches which can be used for to develop a fair predictive model while including race as a predictor. Section 3 provide a discussion on how a fair predictive model can be developed while taking race into account.

## 3. FAIRNESS-AWARE MACHINE LEARNING

These methods usually fall into one of three categories: pre-processing, in-processing, and post-processing. Categorization of these methods depends on where the focus on correcting for discrimination occurs.

With the pre-processing approach, developers adjust the training data in such a way that there are no longer unexplained disparities between protected and unprotected groups. They then develop the model using standard machine learning algorithms. A simple approach is to remove the sensitive variable and all variables correlated with it from the learning process [35]. One of the issues with this method is the loss of information about output in the data at hand. Moreover, correlated variables can be removed but the correlation is a linear function of dependency. Therefore, as discussed in Radovanović and Ivić [36], there is a chance that the interaction of variables leading to unfairness may not be recognized. In other pre-processing approaches, developers might modify the target variable [37-41], modify the input data [35, 42-45] or modify both the target variable and input data [46]. Modifying the data might seem like an appropriate approach. However, this approach will violate the need for data accuracy as emphasized in general data protection regulation (GDPR) [47]. Therefore, the application of such approaches especially in child welfare settings might be questionable.

Methods that fall under the category of In-processing involve modifying the learning algorithm to maximize both predictive accuracy and fairness. For instance, by modifying the splitting criteria in decision tree learning [48, 49]. Most recent in-processing methods add regularizes to the goal function to control for fairness or enforce fairness constraints during the model learning process to turn it into an optimization problem. Refer to [36, 50-59].

In post-processing approaches, one generates a standard model and then adjusts this model to comply with non-discrimination constraints [48, 60-63]. For instance, by changing the labels of some leaves in a decision tree [48, 61], removing selected rules from the set of discovered decision rules [62] or in general adjusting predictions to be as fair as possible [60, 63, 64].

Even though some methods have already been proposed for each of the above approaches, discrimination prevention remains a relatively unexplored area of research. Typically, each solution is customized for a specific setting and discrimination situation and barely generalizes to other types of variables, or other grounds of discrimination. Non-discrimination regulations often specify the sensitive features or the group of people who must be protected against discrimination in a certain setting. Here, the question is how one can determine whether a predictive model is fair to these groups. Clearly, the answer to this question depends on the notion of fairness one wants to achieve. In the following section, some of these concepts are discussed that may be applicable to child welfare settings.

## 4. NOTIONS OF FAIRNESS

Several different notions of fairness proposed in the last decade. Still, there is no transparent agreement on which definition to apply in each situation. Table 2 presents several notions of fairness that could be applied in child welfare settings, along with their mathematical definitions. Table 1 provides a summary of notations that will be used to present these definitions mathematically.

**Table 1.** Summary of notation employed.

| Symbols | Description |
|---------|-------------|
| $s$ | Sensitive variable for which discrimination should be recognised (For example Race to be the sensitive variable and Māori the protected group, let $s = 1$ represents the protected group (disadvantaged group), and $s = 0$ represents the unprotected group (privileged group). |
| $x$ | Other predictor variables describing an individual. |
| $y$ | The actual outcome as presented in the dataset (with $y = 1$ representing the outcome of interest and $y = 0$ representing the opposite). |
| $r$ | Predicted probability or risk score for a certain outcome $i$, denoted $p(y = i \mid s, x)$. |
| $\hat{y}$ | The predicted value of the label variable ($\hat{y}$ is normally obtained from $r$, e.g., $\hat{y} = 1$ when $r$ is greater than a specified threshold). |

**Table 2.** Definitions of fairness.

| Notion of Fairness | Mathematical Definition | Ref. |
|---|---|---|
| Statistical parity | $p(\hat{y} \mid s = 0) = p(\hat{y} \mid s = 1)$ | [65] |
| Predictive parity | $p(y = 1 \mid \hat{y} = 1, s = 0) = p(y = 1 \mid \hat{y} = 1, s = 1)$ | [66] |
| Predictive equality | $p(\hat{y} = 1 \mid y = 0, s = 0) = p(\hat{y} = 1 \mid y = 0, s = 1)$ | [56] |
| Equal opportunity | $p(\hat{y} = 0 \mid y = 1, s = 0) = p(\hat{y} = 0 \mid y = 1, s = 1)$ or $p(\hat{y} = 1 \mid y = 1, s = 0) = p(\hat{y} = 1 \mid y = 1, s = 1)$ | [60] |
| Equalized odd | $p(\hat{y} = 1 \mid y = i, s = 0) = p(\hat{y} = 1 \mid y = i, s = 1)$, $i \in \{0,1\}$ | [60] |
| Accuracy equity | $p(\hat{y} = y, s = 0) = p(\hat{y} = y, s = 1)$ | [55] |
| Calibration | $p(y = 1 \mid r, s = 0) = p(y = 1 \mid r, s = 1)$ | [66] |

The notions of fairness explained in Table 2 are only a few of the definitions proposed in the literature. These definitions are likely to be more appropriate for measuring discrimination and comparing the fairness of the models developed for use by child welfare agencies. Additionally, they can be used to mitigate the discrimination caused by models. This is done by transforming the appropriate definition of fairness into constraints and enforcing them during the learning process of machine learning algorithms. Section 5 will discuss this in greater detail. Other notions of fairness and their mathematical definitions are surveyed by Verma and Rubin [67].

## 5. A POTENTIAL APPROACH TO MITIGATE BIASED OUTCOMES.

In the child welfare settings, the data used to develop predictive risk models are often extracted directly from the child welfare agency's database systems and include records of interactions with the children and their families [8]. Depending on the authority, however, these datasets may be linked to data collected by the government from other agencies such as public hospitals, birth records, public benefits, criminal justice, education records, and more [9, 31, 68]. Often, the outcome variables are dichotomously coded, reflecting whether each child will experience an adverse outcome, such as maltreatment. Thus, it is a binary classification problem. In such cases, it is possible to improve fairness by enforcing appropriate fairness constraints during the learning phase of the logistic regression algorithm, which will thereby result in a model that can generate predictions that are less discriminatory. This potential approach evolved after the review of the studies referenced here [36, 51-53].

In the fairness-aware machine learning field of research, this method is categorized as an in-processing approach to discrimination prevention and was selected primarily for two reasons. First, logistic regression has been selected as a candidate model in previous studies on child welfare due to its simplicity and interpretability. Additionally, there has been no investigation of constrained classification methods in the context of child welfare. An overview of the machine learning algorithms used in earlier studies can be found in Table 3.

**Table 3.** Methodologies used in earlier studies and the candidate models.

| Study | Ref. | Learning Algorithms | Candidate Model |
|---|---|---|---|
| New Zealand | [69] | Probit Regression | Probit Regression |
| New Zealand | [9] | -Gradient Boosting -DMINE Regression -Neural Network -Partial Least Squares -Full Regression -Stepwise Logistic Regression -Regression with backward elimination -Decision tree -Multilevel model | Stepwise Logistic Regression |
| New Zealand | [70] | -Logistic Regression -Decision Tree -Random Forest -Gradient Boosting | Logistic Regression |
| US Allegheny Family Screening Tool (Version 1) | [71] | -Logistic Regression -Random Forest -Support Vector Machine -XGBoost | Logistic Regression |
| Allegheny Family Screening Tool (Version 2) | [72] | -Logistic Regression -LASSO Regression -Random Forest -XGBoost | LASSO Regression |
| Douglas County Decision Aid | [73] | -LASSO Regularized Logistic Regression -Random Forest -XGBoost | LASSO Regression |

As a baseline for comparison, simple logistic regression and Regularized logistic regression algorithms such as Ridge [74], LASSO regression [75], and Elastic net [76] can be considered primarily for two reasons. First, logistic regression was selected as the candidate model due to its' simplicity in prior New Zealand studies and version one of the Allegheny County Family Screening Tool. Secondly, LASSO was used in the development of version two of Allegheny Family Screening Tool and Douglas County Decision Aid because of its overall performance and accuracy for specific high-risk groups as well as its equal accuracy for black children compared to non-black children [72, 73]. The goal of this approach is to compare the results and determine if constrained logistic regression is better in terms of accuracy and fairness compared to state-of-the-art approaches in child welfare settings.

*5.1. Constrained logistic regression.*

It is the intention to develop a fair predictive model by incorporating constraints derived from appropriate notions of fairness into the learning process of the logistic regression algorithm and turning it into an optimization problem.

$$pr(y=1 \mid x ; w) = h_w(x) = \frac{1}{1+e^{-w^T x}}, \quad (5.1.1)$$

Where x represents the predictor variables, and w the weights associated with the predictor variables. Predictor variables are known as the independent variables used to predict an outcome. A logistic regression algorithm finds the weight associated with these variables in a way that the logistic loss function (negative log-likelihood) is minimized (eq.5.1.2).

$$\text{Minimize } L(y, \widehat{y}) = -\frac{1}{n}\sum_{i=1}^{n}(y_i \log(h_w(x_i)) + (1-y)\log(1-h_w(x_i))), \quad (5.1.2)$$

Here y presents real value and $\widehat{y}$ the predicted value of the outcome variable. To apply this in-processing approach for improving fairness (constrained logistic regression), the loss function in equation (5.1.2) will become the objective of the optimization problem and must be minimized subject to fairness constraints that will be defined below.

To proceed with constrained logistic regression, it is necessary to identify an appropriate definition of fairness that fits the problem at hand. These definitions can be selected based on the fairness assessment of the basic logistic regression model.

As an example, an earlier New Zealand study indicated that Māori children are referred by their model at a higher rate than their share of known maltreatment [9]. Rea and Erasmus [10] also revealed that their model refers fewer New Zealand-European and Pacific Island children to the site than is currently the case. This suggests that predictive models developed for the New Zealand population may have a higher False Positive Rate for Mori children and a higher False Negative Rate for children from other ethnicities. It is therefore expected that the model will have a fair False Positive Rate and a False Negative Rate across various ethnic groups.

Equalized odd notion of fairness is satisfied if protected and non-protected group have equal True Positive Rate and equal False Positive Rate [60]. Given that a model with equal False Negative Rate will also have equal True Positive Rate, satisfying an equalized odd notion of fairness during the learning process is appropriate for this problem [60]. See equation (5.1.3).

$$p(\widehat{y}=1 \mid y=i, s=1) = p(\widehat{y}=1 \mid y=i, s=0) \quad i \in \{0,1\} \quad (5.1.3)$$

Moreover, the prediction model should be fair regarding every value of the outcome variable. Therefore, satisfying both statistical parity and equalized odd notions of fairness during the learning process is more appropriate for this problem. Statistical parity is the most common notion of fairness and it's ensured if both protected and unprotected groups have equal probability of the outcome occurring [65]. The mathematical formulation is:

$$p(\widehat{y} \mid s=1) = p(\widehat{y} \mid s=0) \quad (5.1.4)$$

Since these proportions are not often the same, a constant c must be introduced to control for discrimination. Consequently, the equation in (5.1.4) is transformed into (eq.5.1.5) to get a mathematically more suitable definition by considering the constant c.

$$p(\widehat{y} \mid s=1) - p(\widehat{y} \mid s=0) \geq c, \quad (5.1.5)$$

According to the equation in (5.1.5), the difference between the probabilities for the protected and unprotected groups must be greater than or equal to a constant c. Based on the 80% rule in fairness-aware machine learning literature, this value can be set to -0.2 [77]. The 80% rule states that the selection rate of the protected group should be at least 80% of the selection rate of the non-protected group [78]. Accordingly, the difference in probability of the outcome between the protected group and the unprotected group must therefore be greater than or equal to -0.2. Similarly, for equalized odds, the equation in (5.1.3) is transformed to (eq.5.1.6) to get a mathematically more suitable definition by considering the constant c.

$$p(\widehat{y}=1 \mid y=i, s=1) - p(\widehat{y}=1 \mid y=i, s=0) \geq c, \quad i \in \{0,1\} \quad (5.1.6)$$

To enforce these notions of fairness as constraints in the learning process of the logistic regression, they must be transformed into mathematical functions. For this purpose, the constraint (5.1.7) defined in [53] which applies satisfaction of statistical parity (eq. 5.1.5) can be employed.

$$\frac{1}{n}\sum_{i=1}^{n}((s_i - \bar{s}) * p(y_i \mid x_i, w)) \geq c, \quad (5.1.7)$$

Where n is the number of observations, $s=1$ the protected group, $s=0$ the unprotected group and, $\bar{s}$ the proportion of protected group in the population. Since sensitive variable can take two values, the $(s_i - \bar{s})$ can be either positive for the protected group or negative for the unprotected group. The intensity of discrimination is derived by multiplying these values with predicted probabilities derived from logistic regression. Thus, discrimination is estimated as the sum of the impacts of sensitive variables on predictions. In the event of discrimination, the value of the function would be negative, indicating that the unprotected group is dominant. Therefore, this can be added as a constraint, which must be greater than constant c, where c controls the level of discrimination allowed.

Due to the non-linearity of the predicted score in logistic regression, it is possible to use the intensity part of the function ($w^T x_i$) instead of the probability of the outcome, as it is perfectly correlated with it (eq.5.1.8) [51].

$$\frac{1}{n}\sum_{i=1}^{n}((s_i - \bar{s}) * w^T x_i) \geq c, \quad (5.1.8)$$

Using the constraint (eq.5.1.8) will lead both groups to have a similar predictive value of the output on average. But the results can still be unfair. Therefore, an additional constraint that enforces equalized odds is required (e.q.5.1.6) [51]. This constraint is also transformed into a mathematical function (eq. 5.1.9).

$$\frac{1}{n}\sum_{i=1}^{n}((s_i - \bar{s})(y_i-\bar{y})*w^T x_i) \geq c \quad (5.1.9)$$

Where y represents the true outcome and $\bar{y}$ average value of true outcomes. In addition to the statistical parity constraint (eq.5.1.8), this constraint (eq.5.1.9) adds another term ($y_i-\bar{y}$) that enforces that all predicted probabilities between groups remain the same.

Based on the above theories, the problem of learning a fair algorithm in terms of statistical parity and equalized odd can be modified to an optimization problem (eq. 5.1.10) where the logistic regression loss function is minimized subject to constraints (eq. 5.1.8) and (eq. 5.1.9). The optimization problem is:

*Minimize* $L(y, \widehat{y}) = -\frac{1}{n}\sum_{i=1}^{n}(y_i \log(h_w(x)) + (1-y)\log(1-h_w(x)))$,

*where* $h_w(x) = \frac{1}{1+e^{-w^T x}}$,

*subject to* $\frac{1}{n}\sum_{i=1}^{n}((s_i - \bar{s})*w^T x_i) \geq c$, and

$$\frac{1}{n}\sum_{i=1}^{n}((s_i - \bar{s})(y_i-\bar{y})*w^T x_i) \geq c. \quad (5.1.10)$$

This constrained logistic regression (e.q.5.1.10) can be learned using sequential quadratic programming since the objective function is convex and constraints are linear.

*5.2. Selection of the candidate model*

Model selection depends on whether there is a trade-off between concerns about racial bias in the use of the model and loss of accuracy regarding these outcomes. It is therefore necessary to compare methodologies across a series of fairness and accuracy metrics when deciding which method to use. The measures of fairness for this problem are statistical parity and equality of opportunity for a specific class. The statistical parity can be calculated by equation (5.2.1) and equality of opportunity by equation (5.2.2).

*Statistical parity* = $\min\left(\frac{P(\widehat{y}|s=1)}{P(\widehat{y}|s=1)}, \frac{P(\widehat{y}|s=0)}{P(\widehat{y}|s=1)}\right)$,    (5.2.1)

*Equality of Opportunity* = $\min\left(\frac{P(\widehat{y}|s=1,y=i)}{P(\widehat{y}|s=1,y=i)}, \frac{P(\widehat{y}|s=0,y=i)}{P(\widehat{y}|s=1,y=i)}\right)$,
$i \in \{0,1\}$    (5.2.2)

Statistical parity or equality of opportunity must equal one to claim that discrimination does not exist. This would indicate that there are equal opportunities for every group in the dataset. The response to equalized odd can be determined by measuring the equality of opportunity for each class. Therefore, if each class has equal opportunity, equalized odd has been met. However, statistical parity or equality of opportunity equal to 1 is not expected and the models can be compared based on these fairness measures for the selection of the candidate model.

**Discussion and Future Work**

The combination of clinical judgment and predictive tools that use administrative data can assist social workers to predict which children might be at risk, and whether and when authorities should intervene. However, the use of predictive tools in child welfare settings is critical. A predictive model will inevitably make errors. It might identify as low risk some children who go on to experience abuse or neglect (False Negative) and might identify as high risk some children who do not (False Positive). When models are used to classify children based on their risk, these two types of error could lead to different potential harms: A False Positive may lead to unnecessary intervention, even to family separation. At the same time, a False Negative could lead the agency to fail to intervene when it should have. On the other hand, specifically in the New Zealand care and protection system, the over-representation of Māori might be intensified by predictive models. If the data exaggerates the risk, then its' use in decision-making has the potential to feed a cycle of bias that leads to these groups (Māori) being disadvantaged or discriminated against, so that bias or discrimination mitigation should be a goal of the predictive modelling process in child welfare settings. Algorithms are used for decision support, and clarifications are required many times before public concerns and lack of trust lead to unnecessarily restrictive regulatory actions against machine learning.

Our work provides an initial review of possible bias mitigation approaches in the process of developing predictive risk models for use by the New Zealand care and protection system, still developed up to the initial stages. The main objective of this research is to carry out systematic inquiries to examine the facts stated over the potential risk of discrimination associated with the process of developing predictive risk models using machine learning algorithms in New Zealand with a specific focus on Māori population. This study is expected with an immense benefit to the research community, child welfare systems, and children. Firstly, it aims to provide practical solutions and awareness measures to the concerns raised regarding the potential discriminatory effects of using predictive risk modeling within the New Zealand care and protection system. Mitigating the concerns might help the government to progress toward an implementation phase. Researchers and policymakers will understand the potential risk of discrimination in government use of algorithms in decision-making. Secondly, to improve the ability of child protection staff to make more efficient and consistent decisions and will assist them to avoid unnecessary investigations, which are costly for the system, and troublesome for families of racial minority groups such as Māori. This will reduce the number of unnecessary investigations undertaken and better identification of those children and families who are a high priority for services. Finally, it will have an impact on the lives of children who are at risk of maltreatment by identifying their risk score more accurately and preventing severe future outcomes.

The methodology is proposed based on the literature review and the description of datasets intended to be used in future work. Future work will make use of de-identified data from the Integrated Data Infrastructure (IDI) managed by Statistics New Zealand (StatsNZ) to develop a research dataset. An introduction to StatsNZ is provided in the next section.

Although statistical parity and equalized odd notions of fairness seem to be suitable for the problem mentioned in previous New Zealand studies. Still, determining a satisfactory notion of fairness would depend on the research dataset and the assessment of the basic logistic regression model trained on the dataset. For example, the researchers who developed Allegheny Family Screening Tool and Douglas County Decision Aide assessed the fairness of their model in terms of calibration and accuracy equity but they did not try to control it during the machine learning process [28, 73].

Future work is currently in progress. We explore the inclusions of calibration or accuracy equity in form of constraint into the classification model learning process. For accuracy equity, a model satisfies this definition if both protected and unprotected groups have equal prediction accuracy, usually determined by the AUC [55]. The definition of accuracy equity assumes that True Negatives are as desirable as True Positives. A model satisfies the calibration notion if, for any predicted probability score (r), subjects in both protected and unprotected groups have an equal probability of truly belonging to the positive class [79]. This definition is similar to statistical parity except that it considers the fraction of correct positive predictions for any value of the predicted probability 'r' (Table 1). Moreover, the machine learning algorithms do not have to be limited to constrained or regularized logistic regression. Other classification machine learning algorithms such as Random Forest or Support Vector Machine can also be considered, and their results compared.

**Statistics New Zealand (StatsNZ-Tatauranga Aotearoa)**

Statistics New Zealand branded as StatsNZ is a public service department of New Zealand. StatsNZ collects information related to the economy, population, and society of New Zealand from people and organizations through censuses and surveys. They use this information to publish insights and data about New Zealand and support others to use the data [80].

**StatsNZ Integrated Data Infrastructure (IDI)**

The Integrated Data Infrastructure (IDI) is a large research database that contains microdata about people and households from a range of New Zealand government agencies, StatsNZ surveys, and non-government organizations. StatsNZ collects data from different sources and links it together to create integrated data. StatsNZ grants access to the data on a case-by-case basis where a research project meets access criteria and is in the public interest. The data sets have had personal identifiers removed or encrypted so that the risk of disclosure of personal information will be reduced. There is an encrypted identifier for each identity in the IDI that is common across all datasets. This will allow researchers to link variables from multiple sources to gain system-wide insights. Data is bound by the Statistics Act 1975 and the Privacy Act 1993 to protect the identities of people in the data they hold. To protect the privacy of individuals further, data can only be accessed through a secure virtual environment known as the Data Lab, and only in research facilities approved by Stats NZ. Moreover, StatsNZ checks all outputs before they can be released from the Data Lab, to ensure information does not identify individuals. Results that could potentially identify individuals will not be released [81].

*References*

1. Chang, L., *Applying data mining to predict college admissions yield: A case study.* New Directions for Institutional Research, 2006. **131**: p. 53-68.
2. Huang, C.-L., M.-C. Chen, and C.-J. Wang, *Credit scoring with a data mining approach based on support vector machines.* Expert systems with applications, 2007. **33**(4): p. 847-856.
3. Wallace, E., et al., *Risk prediction models to predict emergency hospital admission in community-dwelling adults: a systematic review.* Medical care, 2014. **52**(8): p. 751.
4. Eckerd, *Rapid Safety Feedback: Blue Ribbon Commission on Child Protection.* 2014.
5. Vaithianathan, R., et al., *Allegheny family screening tool: Methodology, version 2*, in *Center for Social Data Analytics*. 2019. p. 1-22.
6. Vaithianathan, R., et al., *Implementing a Child Welfare Decision Aide in Douglas County: Methodology Report*, in *Centre for Social Data Analytics, Auckland*. 2019.
7. Ministry of Social Development, *The White Paper for Vulnerable Children*. 2012, Ministry of Social Development Wellington, New Zealand.
8. Vaithianathan, R., et al., *Children in the public benefit system at risk of maltreatment: Identification via predictive modeling.* American journal of preventive medicine, 2013. **45**(3): p. 354-359.
9. Wilson, M.L., et al., *Predictive modeling: potential application in prevention services.* American journal of preventive medicine, 2015. **48**(5): p. 509-519.
10. Rea, D. and R. Erasmus, *Report of the enhancing intake decision-making project*. 2017, Wellington, New Zealand: New Zealand Government.
11. Ministry of Social Development, *The feasibility of using predictive risk modelling to identify new-born children who are high priority for preventive services—companion technical report*. 2014, Ministry of Social Development Wellington, New Zealand.
12. Dare, T., *Predictive risk modelling and child maltreatment: An ethical review*, in *Auckland, New Zealand, University of Auckland*. 2013.
13. Blank, A., et al., *Ethical issues for Māori in predictive risk modelling to identify new-born children who are at high risk of future maltreatment*. 2015.
14. Drake, B., et al., *A practical framework for considering the use of predictive risk modeling in child welfare.* The ANNALS of the American Academy of Political and Social Science, 2020. **692**(1): p. 162-181.
15. Glaberson, S.K., *Coding over the cracks: predictive analytics and child protection.* Fordham Urb. LJ, 2019. **46**: p. 307.
16. Keddell, E., *Algorithmic justice in child protection: Statistical fairness, social justice and the implications for practice.* Social Sciences, 2019. **8**(10): p. 281.
17. Barocas, S. and A.D.J.C.L.R. Selbst, *Big data's disparate impact.* 2016. **104**: p. 671.
18. Larson, J., et al. *How We Analyzed the COMPAS Recidivism Algorithm*. 2016 [cited 2022 12 September 2022]; Available from: https://www.propublica.org/article/how-we-analyzed-the-compas-recidivism-algorithm.
19. Brantingham, P.J., M. Valasik, and G.O. Mohler, *Does predictive policing lead to biased arrests? Results from a randomized controlled trial.* Statistics and public policy, 2018. **5**(1): p. 1-6.
20. Buolamwini, J. and T. Gebru. *Gender shades: Intersectional accuracy disparities in commercial gender classification*. in *Conference on fairness, accountability and transparency*. 2018. PMLR.
21. Howard, A. and J. Borenstein, *The ugly truth about ourselves and our robot creations: the problem of bias and social inequity.* Science and engineering ethics, 2018. **24**(5): p. 1521-1536.
22. Ruggieri, S., D. Pedreschi, and F. Turini, *Data mining for discrimination discovery.* ACM Transactions on Knowledge Discovery from Data (TKDD), 2010. **4**(2): p. 1-40.

23. Gavighan, C., et al., *Government use of artificial intelligence in New Zealand*. 2019: The New Zealand Law Foundation.
24. Cuccaro-Alamin, S., et al., *Risk assessment and decision making in child protective services: Predictive risk modeling in context*. 2017. **79**: p. 291-298.
25. Eckerd Connects. *Eckerd Rapid Safety Feedback® Highlighted in National Report of Commission to Eliminate Child Abuse and Neglect Fatalities*. 2016 [cited 2022 15 September 2022]; Available from: https://eckerd.org/eckerd-rapid-safety-feedback-highlighted-national-report-commission-eliminate-child-abuse-neglect-fatalities/.
26. Edelman, B.G. and M.J.H.B.S.N.U.W.P. Luca, *Digital discrimination: The case of Airbnb. com.* 2014(14-054).
27. Žliobaitė, I., *Measuring discrimination in algorithmic decision making*. Data Mining and Knowledge Discovery, 2017. **31**(4): p. 1060-1089.
28. Chouldechova, A., et al. *A case study of algorithm-assisted decision making in child maltreatment hotline screening decisions*. in *Conference on Fairness, Accountability and Transparency*. 2018. PMLR.
29. Calders, T. and I. Žliobaitė, *Why unbiased computational processes can lead to discriminative decision procedures*, in *Discrimination and privacy in the information society*. 2013, Springer. p. 43-57.
30. Oranga Tamariki, *Factors Associated with Disparities Experienced by Tamariki Māori in the Care and Protection System*. 2020: Wellington, New Zealand.
31. Vaithianathan, R., et al., *Developing predictive models to support child maltreatment hotline screening decisions: Allegheny County methodology and implementation.* 2017.
32. Lum, K. and J. Johndrow, *A statistical framework for fair predictive algorithms.* arXiv preprint arXiv:1610.08077, 2016.
33. Zliobaite, I., *A survey on measuring indirect discrimination in machine learning.* arXiv preprint arXiv:1511.00148, 2015.
34. Zliobaite, I., *Fairness-aware machine learning: a perspective.* arXiv preprint arXiv:1708.00754, 2017.
35. Calders, T., F. Kamiran, and M. Pechenizkiy. *Building classifiers with independency constraints*. in *2009 IEEE International Conference on Data Mining Workshops*. 2009. IEEE.
36. Radovanović, S. and M. Ivić, *Enabling Equal Opportunity in Logistic Regression Algorithm.* Management: Journal of Sustainable Business and Management Solutions in Emerging Economies, 2021.
37. Kamiran, F. and T. Calders. *Classifying without discriminating*. in *2009 2nd International Conference on Computer, Control and Communication*. 2009. IEEE.
38. Luong, B.T., S. Ruggieri, and F. Turini. *k-NN as an implementation of situation testing for discrimination discovery and prevention*. in *Proceedings of the 17th ACM SIGKDD international conference on Knowledge discovery and data mining*. 2011.
39. Mancuhan, K. and C. Clifton. *Discriminatory decision policy aware classification*. in *2012 IEEE 12th International Conference on Data Mining Workshops*. 2012. IEEE.
40. Mancuhan, K. and C. Clifton, *Combating discrimination using bayesian networks.* Artificial intelligence and law, 2014. **22**(2): p. 211-238.
41. Fish, B., J. Kun, and A.D. Lelkes. *Fair boosting: a case study*. in *Workshop on Fairness, Accountability, and Transparency in Machine Learning*. 2015. Citeseer.
42. Lum, K. and J.J.a.p.a. Johndrow, *A statistical framework for fair predictive algorithms.* 2016.
43. Feldman, M., et al. *Certifying and removing disparate impact*. in *proceedings of the 21th ACM SIGKDD international conference on knowledge discovery and data mining*. 2015.
44. Johndrow, J.E. and K.J.T.A.o.A.S. Lum, *An algorithm for removing sensitive information: application to race-independent recidivism prediction.* 2019. **13**(1): p. 189-220.
45. Kamiran, F. and T. Calders. *Classification with no discrimination by preferential sampling*. in *Proc. 19th Machine Learning Conf. Belgium and The Netherlands*. 2010. Citeseer.
46. Hajian, S. and J. Domingo-Ferrer, *A methodology for direct and indirect discrimination prevention in data mining.* IEEE transactions on knowledge and data engineering, 2012. **25**(7): p. 1445-1459.
47. Regulation, P., *General data protection regulation.* Intouch, 2018. **25**.
48. Kamiran, F., T. Calders, and M. Pechenizkiy. *Discrimination aware decision tree learning*. in *2010 IEEE International Conference on Data Mining*. 2010. IEEE.
49. Kamishima, T., et al. *Fairness-aware classifier with prejudice remover regularizer*. in *Joint European conference on machine learning and knowledge discovery in databases*. 2012. Springer.
50. Calders, T., et al. *Controlling attribute effect in linear regression*. in *2013 IEEE 13th international conference on data mining*. 2013. IEEE.
51. Radovanović, S., et al. *Enforcing fairness in logistic regression algorithm*. in *2020 International Conference on INnovations in Intelligent SysTems and Applications (INISTA)*. 2020. IEEE.
52. Zafar, M.B., et al. *Fairness constraints: Mechanisms for fair classification*. in *Artificial Intelligence and Statistics*. 2017. PMLR.
53. Zafar, M.B., et al., *Fairness Constraints: A Flexible Approach for Fair Classification.* J. Mach. Learn. Res., 2019. **20**(75): p. 1-42.
54. Agarwal, A., et al. *A reductions approach to fair classification*. in *International Conference on Machine Learning*. 2018. PMLR.
55. Berk, R., et al., *Fairness in criminal justice risk assessments: The state of the art.* Sociological Methods & Research, 2021. **50**(1): p. 3-44.
56. Corbett-Davies, S., et al. *Algorithmic decision making and the cost of fairness*. in *Proceedings of the 23rd acm sigkdd international conference on knowledge discovery and data mining*. 2017.
57. Hu, L. and Y. Chen. *Fair classification and social welfare*. in *Proceedings of the 2020 Conference on Fairness, Accountability, and Transparency*. 2020.
58. Johnson, K.D., D.P. Foster, and R.A. Stine, *Impartial predictive modeling: Ensuring fairness in arbitrary models.* arXiv preprint arXiv:1608.00528, 2016.
59. Nabi, R. and I. Shpitser. *Fair inference on outcomes*. in *Proceedings of the AAAI Conference on Artificial Intelligence*. 2018.
60. Hardt, M., E. Price, and N. Srebro, *Equality of opportunity in supervised learning.* arXiv preprint arXiv:1610.02413, 2016.

61. Calders, T. and S. Verwer, *Three naive bayes approaches for discrimination-free classification.* Data Mining and Knowledge Discovery, 2010. **21**(2): p. 277-292.

62. Hajian, S., et al. *Injecting discrimination and privacy awareness into pattern discovery.* in *2012 IEEE 12th International Conference on Data Mining Workshops*. 2012. IEEE.

63. Wu, Y. and X. Wu. *Using loglinear model for discrimination discovery and prevention.* in *2016 IEEE International Conference on Data Science and Advanced Analytics (DSAA)*. 2016. IEEE.

64. Chzhen, E., et al., *Leveraging labeled and unlabeled data for consistent fair binary classification.* arXiv preprint arXiv:1906.05082, 2019.

65. Dwork, C., et al. *Fairness through awareness.* in *Proceedings of the 3rd innovations in theoretical computer science conference*. 2012.

66. Chouldechova, A. and M. G'Sell, *Fairer and more accurate, but for whom?* arXiv preprint arXiv:1707.00046, 2017.

67. Verma, S. and J. Rubin. *Fairness definitions explained.* in *2018 ieee/acm international workshop on software fairness (fairware)*. 2018. IEEE.

68. Vaithianathan, R., et al., *Allegheny family screening tool: Methodology, version 2.* 2019: p. 1-22.

69. Vaithianathan, R., et al., *Children in the public benefit system at risk of maltreatment: Identification via predictive modeling.* 2013. **45**(3): p. 354-359.

70. Rea, D. and R. Erasmus, *Report of the enhancing intake decision-making project.* 2017, Wellington, New Zealand: New Zealand Government, .

71. Vaithianathan, R., et al., *Developing predictive models to support child maltreatment hotline screening decisions: Allegheny County methodology and implementation.* Center for Social data Analytics, 2017.

72. Vaithianathan, R., et al., *Allegheny family screening tool: Methodology, version 2.* Center for Social Data Analytics, 2019: p. 1-22.

73. Vaithianathan, R., et al., *Implementing a Child Welfare Decision Aide in Douglas County.* 2019.

74. Hoerl, A.E. and R.W. Kennard, *Ridge regression: Biased estimation for nonorthogonal problems.* Technometrics, 1970. **12**(1): p. 55-67.

75. Tibshirani, R., *Regression shrinkage and selection via the lasso.* Journal of the Royal Statistical Society: Series B (Methodological), 1996. **58**(1): p. 267-288.

76. Zou, H. and T. Hastie, *Regularization and variable selection via the elastic net.* Journal of the royal statistical society: series B (statistical methodology), 2005. **67**(2): p. 301-320.

77. Feldman, M., *Computational fairness: Preventing machine-learned discrimination.* 2015.

78. Biddle, D., *Adverse impact and test validation: A practitioner's guide to valid and defensible employment testing.* 2017: Routledge.

79. Chouldechova, A., *Fair prediction with disparate impact: A study of bias in recidivism prediction instruments.* Big data, 2017. **5**(2): p. 153-163.

80. StatsNZ. *StatsNZ–Tatauranga Aotearoa*. 2022 [cited 2022 27 April 2022]; Available from: https://www.stats.govt.nz/.

81. StatsNZ. *Integrated Data Infrastructure*. 2022 [cited 2022 25 April 2022]; Available from: https://www.stats.govt.nz/integrated-data/integrated-data-infrastructure/.